\documentstyle[aps,prb]{revtex}         

\begin{document}

\title{Diagrammatic method for investigating universal behavior of impurity
systems} 
\author{Kurt Fischer}
\address{Max-Planck-Institut f\"ur Physik komplexer Systeme, Bayreuther Strasse
40, 01187 Dresden, Germany}
\date{November 18th}
\maketitle

\begin{abstract}
The universal behavior of magnetic impurities in a metal is proved with the
help of skeleton diagrams. 
The energy scales are derived from the structure of the skeleton diagrams. 
A minimal set of skeleton diagrams is sorted out that scales exactly.  
For example, the non-crossing approximation for the Anderson impurity model can
describe the crossover phenomenon. 
The universal Wilson-number is calculated within the non-crossing
approximation.  
The method allows for an assessment of various approximations for impurity
Hamiltonians.    
\end{abstract}


\section{Introduction}
\label{Introduction}

Magnetic impurities in metals show universal behavior at low
energies~\cite{review-Gruener/Zawadowski-Kondo-1}. 
The Hamiltonian of such systems consists of at least one conduction band
of width $D$ to which the impurity couples~\cite{Buch-Hewson}. 
At temperatures $k_B T\ll D$, the scaled observable is independent of details
of the host system such as its band structure. 

A long standing question in the physics of such systems is:
How can the observed universal behavior directly be established from the
original model Hamiltonian such as the Anderson-impurity
model~\cite{Buch-Hewson}, 
together with a reasonable accurate description of observables? 
Hitherto the original Hamiltonian has been replaced by another one which is
more accessible: 

(a) Within the Bethe ansatz the impurity part of the system's thermodynamics
can be derived. 
However, the original model has to be replaced by one with linear dispersion. 
The spectrum of eigenvalues is cut off at
$D^\prime$ which is in general {\it not} identical with the band width $D$ of
the metallic host in the original
model~\cite{review-Andrei/Lowenstein,review-Tsvelick/Wiegmann}.
A relation between $D$ and $D^\prime$ has so far been established only for the
Anderson impurity model~\cite{Rasul/Hewson}.

(b) Within the numerical renormalization group~\cite{review-Wilson}, the
impurity is coupled to a half-infinite chain with hopping matrix elements
vanishing as $  \Lambda^{-n} , n = 1,2,\dots, \Lambda >1$. 
In the limit of $\Lambda  =   1$ one would recover the original
model.  
However, the numerical results have to be extrapolated to that limit because
the length of the chain to be diagonalized numerically would eventually become
too large~\cite{Buch-Hewson,review-Wilson}.      

Hence both methods do not prove that the observables of the original
model behave universally.
On the other hand, there are the diagrammatic approaches to the impurity
problem~\cite{Keiter/Kimball}. 
They allow us to construct approximations for all observables of the original
model.  
The Dyson equation
\[
R^{-1}(z) = R_0^{-1}(z) -\Sigma(z) 
\]
is invariant under a certain rescaling of the propagators $R$ and $R_0$,
self-energies $\Sigma$, and coupling constants. 
With the help of the diagrammatic renormalization group, perturbative results
for the propagators, self-energies, and vertex parts are then fitted to the
Dyson equation. 
In that way, scaling laws for the propagators result, from which the
universal behavior in the perturbative high-energy regime of
the model~\cite{Abrikosov/Migdal} follows.
This method amounts to summing a certain subclass of diagrams, with {\it naked}
propagators. 
However, the procedure breaks down when perturbation theory fails.

For the nonperturbative region {\it dressed} propagators are necessary.
This requires the use of skeleton diagrams, otherwise the definition of a
self-energy itself would become ambiguous. 
However, the scaling of the dressed propagators is unknown, precisely
because beforehand the Dyson equation would have to be solved.
 
In this paper, this difficulty is overcome by utilizing the variational
principle of Luttinger and Ward~\cite{Luttinger/Ward},
Kuramoto~\cite{KuramotoI}, and Baym~\cite{Baym}, 
by which any observable can be expressed in terms of skeleton diagrams.
This is exemplified at the Anderson-impurity Hamiltonian.

It turns out that the skeleton diagrams of the second order are already
sufficient to calculate the exact energy scales.
This proves universal behavior for this model.
The second-order skeleton diagrams are therefore the minimal class of
diagrams which have to be summed, in order to describe
the crossover.

\section{Hamiltonian and diagram technique}
\label{Hamiltonian and diagram technique}

As the standard model describing magnetic impurities in metals,
the Anderson-impurity Hamiltonian~\cite{Buch-Hewson} is considered, 
with a half-filled conduction band of constant density of states
$\rho$ and infinite Coulomb repulsion at the impurity site:
\begin{eqnarray}\label{Andersonmodell-H0-cutoff}
H   &=& H_c + H_f + H_1 ,                       \nonumber \\
H_c &=& \sum_{|\epsilon_p| \leq D,m} ,
        \epsilon_p c_{pm}^+ c_{pm}^{\phantom{+}}        \nonumber \\
H_f &=& \epsilon_f \sum_m f_m^+f_m^{\phantom{+}}        \nonumber \\
H_1 &=& \frac{V}{\sqrt{N}} \sum_{p,m} ,
        \left( c_{pm}^+ b^+ f_m^{\phantom{+}} + H.c. \right) . 
\end{eqnarray}
$c_{pm}^+$ creates a conduction electron with internal quantum number $m = 1
\dots N$, momentum $p$, and energy $\epsilon_p$ which is cut off at $\pm D$.   
$|  m  \rangle = f_m^+ | \text{vac} \rangle$ denotes a magnetic
configuration of the impurity and $|  0  \rangle = b^+ |  \text{vac} 
\rangle$ the non-magnetic one, their energy-difference being $\epsilon_f$.
$f_m^+$ is a fermionic operator and $b^+$ the Coleman boson~\cite{Coleman}.
$F_m^+ = f_m^+b$ creates an electron at the impurity site. 
Double occupancy at the impurity site is suppressed by imposing the constraint
$n_b  +  n_f  = 1$.  
The impurity hybridization with the conduction band is proportional to $V$. 
The Boltzmann constant is set to unity so that temperature is
measured in units of energy.
 
In order to simplify the subsequent derivations, the conduction band of
model~(\ref{Andersonmodell-H0-cutoff}) is assumed to have a constant density
of states $\rho$ with a sharp cutoff at energies $\pm D$, as the other
approaches to scaling
do~\cite{Buch-Hewson,review-Andrei/Lowenstein,Abrikosov/Migdal}.    
Universal behavior should not depend on this
assumption~\cite{review-Andrei/Lowenstein}. 
In Sec.~\ref{How much does scaling depend on band structure?} it will be shown
that the energy scales of the system indeed do not change as long as the
density of states is finite at the Fermi energy, and is sufficient
structureless to have only one energy scale $D$. 

The principal object of concern is the resolvent
$R_f(z) = \langle  Q(z) \rangle_c$ where $Q$ is defined as 
\[
Q(z) = 1/(z - L_c - H_f - H_1)  .
\]
Here the superoperator $L_c$ acts on an operator $X$ of the Hilbert space as
$L_c X = \left[ H_c,X \right]$, and $\langle \rangle_c$ indicates the
thermodynamic average with respect to $H_c$.  
The propagators for the occupied and unoccupied ionic configurations are 
\begin{eqnarray*}
R_m(z) &:=& \langle m| R_f(z) |m\rangle , \nonumber \\
R_0(z) &:=& \langle 0| R_f(z) |0\rangle  .
\end{eqnarray*}
With the help of the identity~\cite{Buch-Fulde}
\begin{equation}\label{identity}
e^{-\beta H} =  e^{-\beta (L_c +H_f+H_1)} e^{-\beta H_c} ,
\end{equation}
the impurity part $Z_f$ of the partition function can then be represented as a
line integral, the path of integration encircling all poles of the integrand: 
\begin{equation}\label{Z-representation}
Z_f := \frac{ \text{Tr}_f \text{Tr}_c e^{-\beta H} }{ \text{Tr}_c e^{-\beta
        H_c} }  
= \text{Tr}_f \oint \frac{dz}{2\pi i} e^{-\beta z} R_f(z)  .
\end{equation}
The well-known diagrammatic technique~\cite{review-Bickers} follows if
$Q$ is expanded in a geometric series in $V$, then $L_c$ acting on the
conduction-electron operators in $H_1$ is evaluated to give the
energy denominators 
\[
L_c(c_{pm}^+ c^{\phantom{+}}_{qn} \dots ) = 
(\epsilon_p - \epsilon_q + \cdots)(c_{pm}^+
c^{\phantom{+}}_{qn} \dots )    ,
\]
and finally Wick's theorem applied to evaluate the thermodynamic average with
respect to $H_c$. 
This can be casted in a diagrammatic language.  
The naked propagators and vertices are shown in
Fig.~\ref{Vertices-Anderson}.  
Because the impurity site can alternatively be empty or singly occupied,
every diagram has a spine of alternating $b$ and $f$ propagators. 
Consequently, all diagrams where a conduction-electron propagator would have
a self-energy are excluded~\cite{review-Bickers}, because they do not fulfill
the constraint $n_b  +  n_f  = 1$.
Alternative approaches where this constraint is not exactly enforced will be
discussed in Sec.~\ref{Alternative diagrammatic methods}.

Within the variational principle~\cite{Luttinger/Ward,Baym}, a
functional $\Upsilon$ of the dressed one-particle propagators is defined.
Because the conduction-electron propagators carry no
self-energy, the variational principle has been adapted to the present
diagrammatic technique~\cite{KuramotoI} such that $\Upsilon$ becomes a
functional of the dressed one-particle impurity-propagators only.
At the saddle point with respect to variations of the $R_{0,m}$, the
functional $\Upsilon$ equals $Z_f$, and the Dyson equation holds as a
self-consistency equation~\cite{KuramotoI}
\begin{equation}\label{Dyson}
R_f(z)^{-1} = z - H_f - \Sigma_f(z)  .
\end{equation}
For the impurity-part of the partition function, the functional is given by  
\begin{eqnarray}\label{Upsilon}
\Upsilon &=& \beta \ \text{Tr}_f \oint \frac{dz}{2 \pi i }  e^{-\beta z}   
         \left\{ \sum_n
         \left( 1-\frac{1}{n} \right) \Sigma_f^{(n)}(z)  R_f(z) \right.  
         \nonumber \\
     &&+ \left.  \ln \left[z-H_f - \sum_n \Sigma_f^{(n)}(z) 
         \right] \right\} . 
\end{eqnarray}
Here $\Sigma_f^{(n)}$ denotes all $n$th order self-energy diagrams of $R_f$,
expressed in terms of skeleton diagrams.
Skeleton diagrams are diagrams where all self-energy insertions have been
removed.
Hence a skeleton diagram becomes a functional of the dressed propagators.  

The variational principle can be interpreted in the following fashion: 
If $\Upsilon$ depends on some parameter $\lambda$ such as the hybridization
then, at the saddle point, it depends on $\lambda$ only explicitly, and not
implicitly via the propagators~\cite{KuramotoI}:
\[
\frac{d \Upsilon}{d \lambda}   =
        \frac{\partial\Upsilon}{\partial \lambda}   
+       \frac{\delta \Upsilon}{\delta R_{0,m}} 
        \frac{\partial R_{0,m}}{\partial\lambda}    
=       \frac{\partial \Upsilon}{\partial \lambda}    .
\]      
Explicitly, from Eq.~(\ref{Upsilon}) it follows at the saddle point that
\begin{eqnarray*}
\frac{d Z_f}{d \lambda} &=&
        - \frac{d}{d\lambda} \,
        \beta \ \text{Tr}_f \oint \frac{dz}{2 \pi i }  e^{-\beta z}   
        \sum_n \frac{1}{n} \Sigma_f^{(n)}(z)  R_f(z)    \\
&&      - \frac{d}{d\lambda} \,
        \beta \ \text{Tr}_f \oint \frac{dz}{2 \pi i }  e^{-\beta z}   
        R_f(z) H_f   ,
\end{eqnarray*}
where the $\lambda$ dependence of $R_f$ can be discarded.

In order to study the dependence of $Z_f$ on a parameter $\lambda$, it
therefore suffices to consider all skeleton diagrams of the type
$\Sigma_{0,m}^{(n)} R_{0,m}$. 
A skeleton diagram for $\Sigma_0^{(2)} R_0$, i.e. of second order, is shown in
Fig.~\ref{skeleton-second-order}.   

Approximations fulfilling the variational principle can be generated
by using a subclass, a so-called family of skeleton diagrams~\cite{Grewe1983},
which contains with each skeleton diagram for $\Sigma_{0,m}^{(n)} R_{0,m}$ all
others with cyclic permuted vertices as well. 
For example, all skeleton diagrams of a given order $\Sigma_{0,m}^{(n)}
R_{0,m}$ form such a family.
 
For instance, if only skeleton diagrams of second order are kept in
Eq.~(\ref{Upsilon}), this amounts to summing all diagrams with bare
propagators and noncrossing conduction-electron lines
and is called the NCA~\cite{KuramotoI}.
The self-energies of the NCA, 
$\Sigma_m^{(2)}(z)  =  \Sigma_m^{(2)} [ R_0(z) ]$ and
$\Sigma_0^{(2)}(z)  =  \Sigma_0^{(2)} [ R_1(z),\dots,R_n(z) ]$,
are then ($f$ denotes the Fermi function)
\begin{eqnarray}\label{NCA equations}
\Sigma_m^{(2)}(z) &=& 
\frac{V^2 \rho }{N} \int_{-D}^D f(\epsilon) R_0(z+\epsilon) d\epsilon \, ,\\
\Sigma_0^{(2)}(z) &=& \frac{V^2 \rho }{N} \sum_{m=1}^N
\int_{-D}^D f(\epsilon) R_m(z+\epsilon) d\epsilon  \nonumber  . 
\end{eqnarray}
The functional $\Upsilon^{(2)}$ has then the form
\begin{eqnarray*}
\Upsilon^{(2)} &=& \oint \frac{\beta dz}{2 \pi i }  e^{-\beta z}  \Bigg\{ 
         \frac{1}{2} \Sigma_0^{(2)}(z)  R_0(z) 
        + \frac{1}{2}  \sum_m \Sigma_m^{(2)}(z) R_m(z) \\
&&      + \ln \left[z- \Sigma_0^{(2)}(z)  \right] 
        + \sum_m \ln \left[z- \epsilon_f - \Sigma_m^{(2)}(z) \right] \Bigg\} . 
\end{eqnarray*}
To construct a variational principle for other observables the Hamiltonian has 
to be coupled to external fields suitably chosen~\cite{KuramotoI}. 
This will be discussed in Sec.~\ref{proof of universality}.

\section{energy scales}
\label{energy scales}

\subsection{First scaling equation}
\label{First scaling equation}

$\Upsilon$ as defined in Eq.~(\ref{Upsilon}) depends explicitly on
$\epsilon_f$ only via the term $H_f  =  \epsilon_f \sum_m
f_m^+f_m^{\phantom{+}}$. 
Consequently, one has for the impurity part $F_f= -T  \ln Z_f$ of the free
energy~\cite{review-Bickers}:   
\begin{equation}
\label{F-ef-derivative} 
Z_f \epsilon_f \partial_{\epsilon_f} F_f = 
\mbox{Tr}_f H_f \oint \frac{dz}{2 \pi i } e^{-\beta z} R_f(z)  .
\end{equation}
$\Upsilon$ depends explicitly on $V$ via~\cite{KuramotoI} the prefactor
$V^{2n}$ of the $2n$th order self-energy $\Sigma_f^{(2n)}$:
\begin{eqnarray}
\label{F-V-derivative} 
Z_f \rho V^2 \partial_{\rho V^2} F_f &=&  
\frac{1}{2} \mbox{Tr}_f \oint
\frac{dz}{2 \pi i } e^{-\beta z} \Sigma_f(z)R_f(z)  .
\end{eqnarray}
To determine the $T$ dependence, the internal integration
variables $z$ and $\epsilon$ as in Eqs.~(\ref{Upsilon}), and ~(\ref{NCA
equations}) are replaced by $Tz$ and $T\epsilon$, respectively.   
The variational principle remains unaffected.
$\Upsilon$ depends now explicitly on $T$ via the prefactor $T^n$ of a skeleton
diagram of $2n$th order, the term $Tz$ in the logarithm, and the integration
boundaries as in Eq.~(\ref{NCA equations}) change to $\pm D/T$.  
Therefore
\begin{eqnarray}
\label{F-T-derivative}
Z_f \left(F_f - T \frac{\partial}{\partial T} F_f \right)  &=&  
\frac{1}{2}  \text{Tr}_f \oint \frac{dz}{2 \pi i } e^{-\beta z} 
\Sigma_f(z)R_f(z)       \nonumber  \\
+ Z_f D \frac{\partial}{\partial D} F_f
&+& \text{Tr}_f H_f \oint \frac{dz}{2 \pi i } e^{-\beta z} R_f(z)  
 .
\end{eqnarray}
Inserting Eq.~(\ref{F-ef-derivative}) and Eq.~(\ref{F-V-derivative}) into
Eq.~(\ref{F-T-derivative}) yields
\begin{equation}\label{F-scaling-first}
F_f =   \left( 
        T \frac{\partial}{\partial T} 
        + \rho V^2  \frac{\partial}{\partial \rho V^2}
        + \epsilon_f  \frac{\partial}{\partial \epsilon_f} 
        + D  \frac{\partial}{\partial D} 
    \right)F_f
     . 
\end{equation}
This equation expresses the scaling of $F_f$ with respect to the
energy scales $D$, $\epsilon_f$, and $\rho V^2$:
\[
F(T,\rho,V,\epsilon_f,D) = T f \left( \frac{T}{\rho V^2} , 
                \frac{T}{\epsilon_f} , \frac{T}{D} \right)  .
\] 
This is henceforth called the first scaling equation.

\subsection{Second scaling equation}
\label{Second scaling equation}

The central issue of this work is to prove and describe the universal behavior
of impurity systems. 
To show the universal behavior for $F_f$, the functional $\Upsilon$ is examined
for large but finite $D$,
that is, when $D$ becomes larger than all other energy scales of the system,
the so-called universal limit~\cite{review-Andrei/Lowenstein}.

$\Upsilon$ depends explicitly on the cutoff $D$ only via the integration
boundaries $\pm D$ of the integration over the conduction-electron energies, as
can be seen in Eq.~(\ref{NCA equations}) for the second-order
skeleton diagrams. 

At first only those second-order self-energies are kept in $\Upsilon$. 
With the help of the spectral densities $\rho_{0,m}$ of $R_{0,m}$ one has
\begin{eqnarray}\label{NCA-D-derivative}
D \frac{\partial}{\partial D} F_f^{(2)} &=& 
\frac{V^2 \rho }{N Z_f} 
        \sum_{\stackrel{\scriptstyle m=1}{\scriptstyle \mu=\pm  1}}^N 
        \int  e^{-\beta\omega} d\omega \   D f(\mu D) 
        [  \rho_m(\omega) \nonumber  \\
&&      \times \, \Re R_0(\omega + \mu D)  + 
        \rho_0(\omega) \, \Re R_m(\omega + \mu D) ]  
\end{eqnarray}
where $\Re R_{0,m}$ denotes the real part of $R_{0,m}$.
In the universal limit, in particular $T/D\to 0$, so that only the terms 
$\propto f(-D) \approx1$ survive. 
Furthermore, in this limit the weighted spectral densities $e^{-\beta\omega}
\rho_{0,m}(\omega)/Z_f$ contribute significantly only for frequencies less then
the impurity part of the ground-state energy $E_0  <  0 $,
because $e^{\beta E_0} Z_f$ tends to 1 for low temperatures.
 From perturbation theory it follows~\cite{KuramotoIII} that $e^{-\beta\omega}
\rho_f(\omega)/Z_f$ vanishes asymptotically as $1/\omega^2$ for large,
negative $\omega$.
Hence it contributes significantly to the integral in
Eq.~(\ref{NCA-D-derivative}) only for
\[
- \sqrt{D} \lesssim \omega \lesssim E_0  .
\]  
In this frequency interval, $\Re R_f ( \omega-D)$ can be replaced by
its bare counterpart $1/(\omega - H_f - D) \approx (-1)/D$:
\[
D \frac{\partial}{\partial D} F_f^{(2)} =  
\frac{- V^2 \rho }{N Z_f} 
        \sum_{m=1}^N \int  e^{-\beta\omega} d\omega 
        \left[ \rho_m(\omega) + \rho_0(\omega) \right]  .
\]
Together with Eq.~(\ref{F-ef-derivative}), the following scaling
relation is obtained: 
\begin{equation}\label{F-scaling-second}
D \frac{\partial}{\partial D} F_f^{(2)} =  
        \rho V^2  \left( 1- 1/N \right)
        \frac{\partial}{\partial \epsilon_f} F_f^{(2)} - \rho V^2  .
\end{equation}
This is called henceforth the second scaling equation.
Next it is shown that in the universal limit all families of higher order
skeleton diagrams of $\Upsilon$ are {\it irrelevant}, by which is meant here
that their contribution to the logarithmic derivative in
Eq. (\ref{F-scaling-second}) vanishes as $O(1/D)$. 
For the proof see Appendix~\ref{Irrelevance of higher-order
skeleton diagrams}, the families of skeleton diagrams being needed to enforce
the variational principle.

The result can be made plausible by casting it into the language of diagrams,
as in Fig.~\ref{skeleton-second-order}.
Differentiating logarithmically with respect to $D$ means removing one curved
conduction-electron line and replacing the internal propagator by its value at
the cutoff $\propto  1/D$.
Therefore this diagram contributes $\propto  D/D$ to the logarithmic
derivative. 

For a diagram of higher order than two such as in Fig.~\ref{skeleton-sixth-order},
there lie under {\it each} conduction-electron line at least three
propagators, because otherwise this diagram would have a self-energy
insertion.  
Hence its contribution to the logarithmic derivative is $\propto  D/D^3$ and
can be neglected for large $D$.

Equation~(\ref{F-scaling-second}) therefore holds as well for the {\it exact}
$F_f$. 
It describes the Haldane scaling~\cite{Haldane-Skalierung} for the
difference $F_f-E_0$, that is, it depends on $D$ and $\epsilon_f$ only via 
\[
(F_f - E_0) ( T, V, \rho, \epsilon_f, D ) = 
(F_f - E_0) ( T, V, \rho, \epsilon_f^*) ,
\]
\[
\epsilon_f^* = \epsilon_f + (1-1/N) \rho V^2 \ln D/ (\rho V^2)  . 
\]
By inserting this into Eq.~(\ref{F-scaling-first}), in the magnetic
limit $-\epsilon_f \gg \rho V^2$ the scaling law 
\begin{equation}\label{F-scaling}
F_f - E_0 = 
T  g(T,V,\epsilon_f,D, \rho) =  T  g( \frac{T}{\Gamma} ,
\frac{T}{T_K} ) 
\end{equation} 
follows, with the quantity $\Gamma  =  \pi \rho V^2/N$ and (up to a numerical
factor) the
Kondo temperature\cite{Buch-Hewson} 
\begin{equation}\label{TK}
T_K =  D \sqrt[N]{ \rho V^2 / D } \exp\left[  \epsilon_f/ \rho V^2 \right]
 .
\end{equation}
This proves the universal behavior for the free energy of the Anderson
Hamiltonian as well as that the energy scales $\Gamma$ and $T_K$ are the exact
ones.  
In particular, it has been shown that the NCA preserves the {\it
exact} energy scales of the system.   
Because the NCA is exact up to orders $V^4$ and $1/N$, 
an inclusion of families of skeleton diagrams of higher order in
Eq.~(\ref{F-scaling}) will slightly change $g$ but {\it not the energy
scales}, and consequently will not alter the approximation qualitatively.
Such an extension of the NCA has been performed~\cite{Anders-Post-NCA} and
within the errors of the numerical calculation the scaling
law~(\ref{F-scaling}) as well as a slight change in the respective function
$g$ have been verified.  

There holds an analogy to the usual diagrammatic
renormalization group, as described in Sec.~\ref{Introduction}. 
It turns out that from a certain order in perturbation theory on, the energy
scales obtained by that method do not change any more in the universal limit, 
while, for an observable the form of the respective function $g$ can still
change. 
However, the point is that this scheme can only be used for high
temperatures $T\gg T_K$, below which perturbation theory breaks down.
Hence a necessary condition for a diagrammatic technique to describe the
universal behavior of impurity systems is that it includes families of
skeleton diagrams {\it including} those of second order.  
This result can straightforwardly be extended to the case of finite magnetic
fields.

\section{scaling with magnetic field}
\label{scaling with magnetic field}

The influence of a magnetic field $h$ on the host metal is of order 
$h/D  \propto  1/D$, and can be neglected.  
Hence, to the Hamiltonian of Eq.~(\ref{Andersonmodell-H0-cutoff}) only 
\begin{equation}\label{magnetic-field-substitution}
H_f \longrightarrow H_f + g \mu_B h \sum_m m \, f_m^+f_m^{\phantom{+}} 
\end{equation}
is added, where $g$ denotes the $g$ factor of the $f$ electron and $\mu_B$ is
the Bohr magneton.  
The functional $\Upsilon$ will explicitly depend on $h$ only via the new term
in $H_f$.
Consequently the first scaling equation~(\ref{F-scaling-first}) changes to  
\begin{eqnarray}\label{F-Skalierung-Magnetfeld}
F &=&   T\frac{\partial}{\partial T} F 
        + \rho V^2\frac{\partial}{\partial \rho V^2} F +
        \epsilon_f\frac{\partial}{\partial \epsilon_f} F \nonumber \\
&&      + g\mu_B h \frac{\partial}{\partial g\mu_B h} F +
        D \frac{\partial}{\partial D}  F  .
\end{eqnarray}
The second scaling equation~(\ref{F-scaling-second}) remains unaffected.
The scaling law for the magnetization 
$M(h)  = - ( \partial / \partial h ) F_f$ therefore reads 
\begin{equation}\label{M-scaling}
M(T,V,\epsilon_f,D,\rho,h) = M \left( \frac{T}{\Gamma} , \frac{T}{T_K} ,
                        \frac{h}{\Gamma} , \frac{h}{T_K} \right)  .
\end{equation}

\section{proof of universality}
\label{proof of universality}

To obtain the scaling of any observable, the Hamiltonian $H$ has to be coupled
to suitable external fields.  
However, only the skeleton diagram of Fig.~\ref{skeleton-second-order} is
relevant for large $D$.
Analogous scaling equations such as Eqs.~(\ref{F-scaling-first})
and~(\ref{F-scaling}) can thus be derived for any other observable, hence
proving universality for the Anderson model. 

As an example, take the $f$ propagator~\cite{review-Bickers} as a function of
imaginary time $\tau$,
\begin{equation}\label{Gf-definition}
G_{fm}(\tau) = - \langle  T_\tau F_m(\tau)F_m^+(0) \rangle  .
\end{equation}
Here $T_\tau$ is the time-ordering operator, and $\langle \rangle$ denotes the
thermodynamic average with respect to $H$.
With the help of $Q(z)$ one can reproduce the well-known integral
representation for $G_{fm}$ which coincides for fermionic Matsubara
frequencies with the Fourier transform of
$G_{fm}(\tau)$~\cite{Keiter/Morandi},    
\[
G_{fm}(\omega) =  \oint e^{ -\beta z} \frac{dz}{ 2\pi i }
\langle  \langle  0|Q(z)|0 \rangle \langle  m | Q(z + \omega) | 
m \rangle \rangle_c  .
\]
To obtain a variational functional for $G_{fm}$, $H$ is coupled to an
auxiliary fermionic field $\Psi$: 
\[
H +  \omega  \Psi^+ \Psi^{\phantom{+}} 
  + V \sqrt{\rho} \lambda ( \Psi^+ b^+ f_m^{\phantom{+}} + H.c. )   .
\]
The $f$ propagator is obtained from all diagrams of second order in $\lambda$
that is, after removing the $\Psi$ propagator~\cite{KuramotoI}. 
A functional $\Upsilon_\Psi$ can be constructed analogously to
Eq.~(\ref{Upsilon}), to give, at the saddle point,
\[
Z_\Psi :=       \frac{ \text{Tr}_f \text{Tr}_c \text{Tr}_\Psi e^{-\beta H} }
                { \text{Tr}_c \text{Tr}_\Psi e^{-\beta (H_c+ H_\Psi)} }  .
\]
$G_f$ follows as
\begin{equation}\label{Gf-aus-erzeugendem-Funktional}
\left. \frac{\partial}{\partial \lambda^2} \right|_{\lambda = 0}  F_\Psi  
=  \rho V^2 f(\omega) G_{fm}(\omega)   .
\end{equation}
The scaling equations for $F_\Psi$, and via
Eq.~(\ref{Gf-aus-erzeugendem-Funktional}) for $G_f$, can be  
established in the same manner as for $F_f$. 
In particular, Eq.~(\ref{F-scaling-first}) now reads
\begin{equation}\label{Gf-first-scaling}
\left(  T \frac{\partial}{\partial T} 
+ \omega \frac{\partial}{\partial \omega} 
+  \epsilon_f \frac{\partial}{\partial \epsilon_f} 
+  \rho V^2 \frac{\partial}{\partial \rho V^2} 
+  D \frac{\partial}{\partial D}  
+ 1  \right)  G_{fm} = 0 .
\end{equation}
In the universal limit of large $D$ again only the {\it skeleton diagrams of
second order} are relevant, and the analog to Eq.~(\ref{F-scaling-second})
holds for $G_f$, too:
\[
D \frac{\partial}{\partial D} G_{fm} =  
        \rho V^2  \left( 1- 1/N \right)
        \frac{\partial}{\partial \epsilon_f} G_{fm}   .
\]  
In the magnetic limit $-\epsilon_f \gg \Gamma$ where 
$\Gamma  \gg  T_K$ it follows that the Abrikosov-Suhl resonance and the
impurity part of the resistivity computed within the NCA scale with the 
exact Kondo temperature.  
This was numerically observed in Ref.~\onlinecite{Bickers/Cox/Wilkins}. 
There, the energy scale was determined as the maximum value of the spectral
density of $G_{fm}(\omega)$ in the magnetic limit of the Anderson model, and
{\it assumed} to be proportional to $T_K$. 
Here this has been proved.

\section{Description of Crossover: Wilson number}
\label{Description of Crossover: Wilson number}

Within this scaling method the crossover phenomenon~\cite{review-Wilson}
can now be described entirely within a diagrammatic approach.
Even for the skeleton diagrams of second order an analytical solution of the
self-consistency equations does not seem possible for finite temperature. 
However, for zero temperature the well-known expressions for the ground-state
energy of the NCA~\cite{KuramotoI,Mueller-Hartmann-NCA} for the Anderson model
can be evaluated analytically (see Appendix~\ref{NCA at zero
temperature}) in the magnetic limit.
In this limit, it follows from Eq.~(\ref{M-scaling}) that for zero
temperature the scaling law
\begin{equation}\label{M-scaling-T=0}
M(V,\epsilon_f,D,\rho,h) = M\left( \frac{h}{T_K} \right) 
\end{equation}
holds, which holds as well for the NCA, as shown above.
For low magnetic fields, the low-field energy scale $T_L$ can be fixed
unambiguously by the static magnetic susceptibility at vanishing magnetic
field~\cite{Buch-Hewson},
\[
\chi(0)= \frac{1}{3} \mu_j^2 / T_L  .
\]
With the help of the analytical NCA results of Appendix~\ref{NCA at zero
temperature},  an {\it analytical} expression for $T_L$ is obtained:
\begin{equation}\label{TL-NCA}
T_L^{\text{NCA}} = T_K / \Gamma(1-1/N) ,
\end{equation}
with $\Gamma$ being the gamma function.
The result coincides up to order $1/N$ with the result $T_L = T_K
\Gamma(1+1/N)$ which is believed to be exact~\cite{Buch-Hewson}.  

For high magnetic fields $g \mu_B h \gg T_K$, the NCA reproduces the
perturbation theory up to $V^4$.
With the definitions $N = 2j +1 $ 
and $J = V^2 / |\epsilon_f|$, for the
impurity part of the magnetization the expansion
\begin{eqnarray*}
\frac{M(h)}{jg\mu_B} 
&=&     1 - \frac{\rho J}{N}+ \frac{(\rho J)^2}{N} \ln \frac{
        g \mu_B h  \ \sqrt[N]{e} }{ 
        D \sqrt[N]{\epsilon_f/D}  }   \\
&&      + \frac{(\rho J)^2}{N} \left[ 
        \frac{2}{N(N-1)} {\displaystyle \sum_{m=1}^{N-1} } m  \ln (m)
        \right]
        + O(J^3)   
\end{eqnarray*}
is obtained.
The exact scaling~(\ref{M-scaling-T=0}) of the NCA renders it possible to
reexpress in the perturbation expansion $J$ in terms of $h/T_K$.
The high-field energy scale $T_H$ is fixed unambiguously~\cite{Buch-Hewson} by
requiring that terms $\propto 1 / \ln^2 \left[ g\mu_B h/T_H \right] $
to be absent in the resulting asymptotic expansion.
One arrives at the well-known asymptotic renormalization-group result for the
magnetization in high magnetic fields 
\begin{eqnarray}\label{M-asymptotic}
\frac{M(h)}{jg\mu_B} 
&=&     1 -  \frac{1}{N} \frac{1}{
        \displaystyle \ln \frac{g\mu_B h}{T_H} } 
        - \frac{1}{N^2} \frac{\displaystyle \ln \ \ln \frac{g\mu_B h}{T_H}  
        }{\displaystyle  \ln^2 \frac{g\mu_B h}{T_H}  } \nonumber \\
&&      + O \left(  \frac{\displaystyle \ln \ \ln \frac{g\mu_B h}{T_H}  }{
        \displaystyle  \ln^3 \frac{g\mu_B h}{T_H}  } 
        \right)  
\end{eqnarray}
and $T_H$ is given by
\begin{equation}\label{TH}
T_H = T_K \exp\left[ 
        - 1/N - \frac{2}{N(N-1)}  \sum_{m=1}^{N-1}  m \ \ln (m) \right] .
\end{equation}
The Wilson ratio $W  =  T_H/T_L$ characterizes the crossover from the
high-field region where the impurity reacts as an asymptotically free magnetic
moment, to the low-energy region where the impurity is screened. 
Hence the Wilson ratio coincides with the exact result up to order $1/N$:
\begin{eqnarray}\label{Wilson-number}
\frac{ W^{\text{NCA}} } { W } 
&=&     \frac{1}{ \Gamma(1-1/N) \Gamma(1+1/N) } \nonumber \\
&=&     1 + O ( 1/N^2 )  .
\end{eqnarray}
By a Schrieffer-Wolff transformation~\cite{review-Bickers}, this
result can be extended to the
Coqblin-Schrieffer model (see Appendix~\ref{Coqblin-Schriefer model}).

\section{How much does scaling depend on band structure?}
\label{How much does scaling depend on band structure?}

The universal behavior of impurity models like the Anderson model should not
depend on details of the host's density of states as long as its band has a
finite density of states at the Fermi energy~\cite{review-Andrei/Lowenstein}.  
This was realized for model~(\ref{Andersonmodell-H0-cutoff}) by assuming a
constant density of states with a sharp cutoff at $\pm D$, as the other
approaches to scaling
do~\cite{Buch-Hewson,review-Andrei/Lowenstein,Abrikosov/Migdal}.
There it is assumed that the energy scales of the system do not change as
long as the density of states is finite at the Fermi energy, and scales as  
$\widetilde{\rho}(\epsilon) = \rho(\epsilon / D )$,
which means that the band is sufficient structureless to have only one
energy scale $D$.

Here this is proved: 
The first scaling equation~(\ref{F-scaling-first}) can be taken over because of
\[
T \frac{\partial}{\partial T} \rho(\epsilon T /D ) = 
- D \frac{\partial}{\partial D} \rho(\epsilon T /D )  . 
\]
For the second scaling equation~(\ref{F-scaling-second}), the term 
\[
Y : = V^2 \int_{-\infty}^\infty
\rho(\epsilon/D) \Lambda_2(\omega + \epsilon) f(\epsilon) d\epsilon
\]
as in Eq.~(\ref{Lambda1-Lambda2}) in Appendix~\ref{Irrelevance of higher-order
skeleton diagrams} is examined, 
which represents an integration over a conduction-electron line. 
$\Lambda_2$ stands for the real part of the $m$ propagators which lie under
this conduction-electron line.   
Differentiating logarithmically with respect to $D$ and substituting
$\epsilon :  =  uD$ yields 
\[
D \frac{\partial}{\partial D} Y = 
(-D) \int_{-\infty}^\infty
u \rho^\prime(u) f(uD) \Lambda_2(\omega + uD) \, du .
\]
The integrand can be neglected outside the interval $-\sqrt{D} \lesssim
\omega \lesssim 0$ as in Sec.~\ref{Second scaling equation}.
If $\rho$ is sufficiently smooth around the Fermi energy, the integrand
contributes significantly only for  
$-1 \lesssim u \lesssim 1/\sqrt{D} $. 
Therefore one can replace  $\Lambda_2(\omega + uD)$ 
by its asymptotic value $1/(-uD)^m$, and the Fermi function by its
$T=0$ values up to terms of order $1/D$.  
$Y$ scales as
\[
D \frac{\partial}{\partial D} Y \propto D^{1-m}  .
\]
Only the skeleton diagrams of second order $(m=1)$ are relevant:
\[
D \frac{\partial}{\partial D} Y \propto
V^2 \int_{-\infty}^0  \rho^\prime(u) du = V^2 \rho(0)  . 
\]
Thus the second scaling equation is still valid and depends on the density of
states only through its value at the Fermi energy.

\subsection*{Magnetic impurity in a superconductor}
\label{Magnetic impurity in a superconductor}

If the density of states of the host has still one energy scale
$D$, but the density at the Fermi surface vanishes according to a
power law, this models magnetic impurities in
superconductors~\cite{Satori/Shiba/Sakai/Shimizu} with gap nodes: 
\begin{eqnarray}\label{rho-vanishing}
\rho(\omega) &=& (1+r) \bar{\rho} \left( \frac{\omega}{D} \right)^r , \\
\int_{-D}^D \rho(\omega) d\omega &=& 2 D \bar{\rho}  . \nonumber 
\end{eqnarray}
This modifies the scaling law~(\ref{F-scaling-second}) accordingly:
Every skeleton diagram depends explicitly on $D$ via the integration
boundaries, which sorts out the NCA diagrams as the only relevant diagrams.
In addition, every skeleton diagram has one factor $D^{-r}$ per loop. 
Hence, the second scaling equation is
\begin{eqnarray}\label{rho-vanishing-scaling-F-second}
D \frac{\partial}{\partial D} F_f 
&=& (1+r)(1-1/N)\bar{\rho} V^2 \frac{\partial}{\partial \epsilon_f} 
        F_f \nonumber \\
&& -    r \bar{\rho} V^2 \frac{\partial}{\partial \bar{\rho} V^2} F_f 
        - (1+r) \bar{\rho} V^2  .
\end{eqnarray}
The first scaling equation remains unchanged. 
In the limit of large $N$, this scaling equation is consistent with the
results of Fradkin~\cite{Withoff/Fradkin},
\begin{equation}\label{TK-Withoff/Fradkin}
T_K = D \left( 1 - \frac{r}{(1+r) J \bar{\rho}} \right)^{1/r}  .
\end{equation}

\section{Alternative diagrammatic methods}
\label{Alternative diagrammatic methods}

What conclusions can be drawn from this theory as to the reliability of
other diagrammatic approaches, especially their ability to describe the
crossover?  

\subsection{NCA and the $1/N$ expansion}
\label{NCA and the 1/N expansion}

If the system has $N$ internal degrees of freedom and exhibits
local Fermi-liquid behavior, the $1/N$ expansion becomes exact in the limit
of large $N$~\cite{Read/Newns-slave-boson}.
Contrary to common belief~\cite{Buch-Hewson,Kuroda/Jin}, the $1/N$ expansion is
{\it not} suited for the perturbative regime of high temperatures or magnetic
fields because it fails to reproduce the diagrammatic 
renormalization-group results. 
That is so because the Kondo temperature itself is a function of $1/N$. 
Hence it is not possible to describe the {\it crossover} within a finite-order
$1/N$ expansion.  

The NCA was considered so far as a ``self-consistent''
$1/N$ expansion~\cite{review-Bickers,Kuroda/Jin}.   
However, this does not explain why the NCA values for the static magnetic
susceptibility relative to their $T = 0$ value agree so well with the
respective renormalization-group results~\cite{Zhang/Lee}, even for $N = 2$.   
In view of Eq.~(\ref{M-scaling}), this now becomes clear.

\subsection{Higher order skeleton diagrams}
\label{Higher order skeleton diagrams}

One may ask whether it is possible to improve the NCA substantially by
incorporating in $\Upsilon$ skeleton diagrams of higher order.  
However, one has to be aware that then the numerical problems in solving the
self-consistency equations become formidable~\cite{Anders-Post-NCA}. 
To date one has not succeeded to calculate $\chi(0)$, and hence the
Wilson number up to order $1/N^2$ by this diagrammatic approach.

\subsection{Non-self-consistent methods}
\label{Non-self-consistent methods}

One way of incorporating higher-order skeleton diagrams in a theory for
impurity systems was put forward by Saso~\cite{Saso} in his $T$-matrix
approach. 
The impurity propagators $R_f$ were calculated within the NCA.
For an observable, these NCA propagators were inserted into the respective
skeleton diagrams of orders higher than 2. 
Such an approach cannot be derived from a variational principal.
Hence it will not correctly describe the universal behavior of the impurity.

\subsection{Coleman's diagram technique}
\label{Coleman's diagram technique}

In Coleman's approach~\cite{Coleman} to the
Hamiltonian~(\ref{Andersonmodell-H0-cutoff}), the constraint $n_f + n_b = 1$ is
enforced by adding $\lambda (n_f+n_b)$ to $H$ and sending $\lambda$ to
infinity; consequently, $H_f$ in Eq.~(\ref{Andersonmodell-H0-cutoff}) changes
to 
\begin{equation}\label{Andersonmodell-Coleman}
H_f = (\epsilon_f+ \lambda) \sum_m f_m^+f_m^{\phantom{+}}       
        + \lambda b^+ b^{\phantom{+}}    . 
\end{equation}
The scaling equations~(\ref{F-scaling-first})
and~(\ref{F-scaling-second}) are rederived with the help of 
this diagram technique, and then Coleman's technique is used to discuss the
conserving slave-boson approach of Kroha {\it et
al}~\cite{Kroha/Hirschfeld/Muttalib/Woelfle}.  

Now that the unperturbed part $H_c+H_f$ is a one-particle Hamiltonian,
the standard Matsubara-perturbation theory can be developed.
The naked propagators from which the diagram technique is built are given by
\begin{eqnarray}\label{naked-propagators-Coleman}
R_m^{(0)}(i\omega_n) 
&=&     1/(i\omega_n - \epsilon_f - \lambda) , \nonumber \\
R_0^{(0)}(i\omega_n) ,
&=&     1/(i\nu_p - \lambda) \nonumber \\
R_{cm}^{(0)}(i\omega_n) 
&=&     \rho \int_{-D}^D
        \frac{1}{i\omega_n - \epsilon} d\epsilon  . 
\end{eqnarray}
Here $i\omega_n  =  2\pi (n + 1/2) T$ and $i\nu_p  =  2\pi n T$ 
are the fermionic and bosonic Matsubara frequencies, respectively.
The vertices are displayed in Fig.~\ref{Vertices-Anderson}.
Again, a dashed line denotes $R_m^{(0)}$, a wavy line $R_0^{(0)}$, and a solid
line represent the propagator of a conduction electron $R_{cm}^{(0)}$.
This propagator already contains the sum over all momenta because of the local
interaction with the impurity. 

Skeleton diagrams can be defined as above as containing no self-energy
insertions. 
The variational principle follows with the help of the functional 
\begin{eqnarray}\label{Upsilon-Coleman}
\lefteqn{
- \frac{ \widetilde{\Upsilon} }{T}  
:=      \sum_{m,n,v} \left( 3 - \frac{2}{v} \right)
        \Sigma_m^{(v)}(i\omega_n) R_m(i\omega_n)        
} \nonumber \\
&&      + \sum_{mn} [ 
        \ln\left( i\omega_n -\epsilon_f -\lambda -\Sigma_m(i\omega_n) \right) 
        - \ln( i\omega_n - \epsilon_f - \lambda ) ]     \nonumber \\
&&      + \sum_{mn} \{ \ln ( \left[R_{cm}^{(0)}\right]^{-1} 
        - \Sigma_{cm} (i\omega_n) )
        - \ln\left[R_{cm}^{(0)}\right]^{-1} \}          \nonumber \\
&&      - \sum_{p,v} \left[ \ln \left( i\nu_p - \lambda - \Sigma_0 (i\nu_p)
        \right)  - \sum_p \ln( i\nu_p - \lambda ) \right]   . 
\end{eqnarray}
The factor $3 - 2/v$ arises because every term $\Sigma^{(v)}R$ contains
$3v/2$ propagators.
It is straightforward to show that $\widetilde{\Upsilon}$ is stationary with
respect to variations of the propagators iff the Dyson equations hold.
By rearranging the internal summation-frequencies one has the following
identity: 
\begin{eqnarray}\label{shifting-identity-Coleman}
\sum_{m,n} \Sigma_m(i\omega_n) R_m (i\omega_n)
&=& \sum_{m,n} \Sigma_{cm}(i\omega_n) R_{cm} (i\omega_n) \nonumber \\
&=& (-1)  \sum_p  \Sigma_0 (i\nu_p) R_0 (i\nu_p)   ,
\end{eqnarray}
the minus sign in the second line coming from the additional
fermion loop~\cite{Coleman}.   

As in Luttinger's original approach~\cite{Luttinger/Ward}, one shows that
$\widetilde{\Upsilon}$ is equal to the difference of the interacting and 
noninteracting free energy $F$:
\begin{equation}\label{Upsilon-Coleman-saddlepoint-value}
\widetilde{\Upsilon} = F(V) - F(0) =: \Delta F  .
\end{equation}
The Hilbert space is a sum of eigenspaces of the number operator 
$n  =  \sum_m f_m^+f_m^{\phantom{+}} + b^+ b^{\phantom{+}} $.
The partition function can then be represented as a sum over these subspaces.
The subspace $n=0$ describes the noninteracting system.
In Coleman's original approach, the physical subspace $n=1$ is projected out by
\begin{equation}\label{F-lambda-projection}
\lim_{\lambda\to\infty} \frac{\partial}{\partial e^{-\beta\lambda}} 
        \frac{-\Delta F}{T} = \lim_{\lambda\to\infty} Z_f(\lambda) = Z_f
          .
\end{equation}
The limit is approached smoothly so that 
\begin{equation}\label{Z_f-smooth}
\lim_{\lambda\to\infty} \frac{\partial}{\partial\lambda} Z_f(\lambda) = 0 
.
\end{equation}

\subsubsection{First scaling equation}
\label{First scaling equation-Coleman}

Analogously to Sec.~\ref{First scaling equation}, the explicit derivatives
with respect to $\rho V^2$, $\epsilon_f$, and $\lambda$ are
\begin{eqnarray}\label{DF-V-derivative} 
\rho V^2 \frac{\partial}{\partial \rho V^2} \frac{\Delta F}{T} 
&=&     \sum_{m,n,v} \Big[ \Sigma_m^{(v)}(i\omega_n) R_m(i\omega_n)\Big] , \\
\label{DF-ef-derivative} 
\frac{\partial}{\partial\epsilon_f} \frac{\Delta F}{T}  
&=&     \sum_{m,n} R_m(i\omega_n)       ,                                \\
\label{DF-lambda-derivative} 
\frac{\partial}{\partial\lambda} \frac{\Delta F}{T}    
&=&     \sum_{m,n} R_m(i\omega_n) - \sum_p R_0(i\nu_p)  .
\end{eqnarray}
To determine the $T$ dependence of $\widetilde{\Upsilon}$, one has to bear in
mind that each Matsubara frequency is proportional to $T$, that each summation
over Matsubara frequencies gives a factor $T$, and that in a diagram of order
$v$ there are $v/2$ such summations.  
Furthermore, $R_{cm}^{(0)}$ depends on $T$ and $D$ only via $D/T$.
Together with Eqs.~(\ref{shifting-identity-Coleman}), (\ref{DF-V-derivative}),
(\ref{DF-ef-derivative}), and~(\ref{DF-lambda-derivative}) it follows that
\begin{equation}\label{DF-scaling-first}
\Delta F =  \left( 
T \frac{\partial}{\partial T} 
+ \epsilon_f\frac{\partial}{\partial \epsilon_f} 
+ \rho V^2\frac{\partial}{\partial \rho V^2}
+ \lambda\frac{\partial}{\partial \lambda} 
+ D\frac{\partial}{\partial D}  \right)  \Delta F .
\end{equation}
By using Eqs.~(\ref{F-lambda-projection}) and~(\ref{Z_f-smooth}), the
analog of the scaling equation~(\ref{F-scaling-first}) follows:
\begin{equation}\label{Z_f-scaling-first}
\left( 
T \frac{\partial}{\partial T} 
+ \epsilon_f\frac{\partial}{\partial \epsilon_f} 
+ \rho V^2\frac{\partial}{\partial \rho V^2}
+ D\frac{\partial}{\partial D} \right) Z_f = 0   . 
\end{equation}

\subsubsection{Second scaling equation}
\label{Second scaling equation-Coleman}

The functional $\widetilde{\Upsilon}$ depends on the cutoff $D$ only via the
integration boundaries $\pm D$ in $R_{cm}^{(0)}$. 
Hence
\begin{eqnarray}\label{DF-D-derivative-preliminary}
D\frac{\partial}{\partial D} \frac{\Delta F}{T} 
&=&     \sum_{mn} \left[ R_{cm}^{(0)}(i\omega_n) \right]^{(-2)} 
          \left( \frac{D}{i\omega_n +D} + \frac{D}{i\omega_n -D} \right)
        \nonumber \\
&&\times [ R_{cm}(i\omega_n) - R_{cm}^{(0)}(i\omega_n) ]  .
\end{eqnarray}
The propagator difference $R_{cm}-R_{cm}^{(0)}$ can be expanded as
\begin{equation}\label{R-series}
Y:=      \left( R_{cm}- R_{cm}^{(0)}\right) 
         \left[ R_{cm}^{(0)} \right]^{(-2)} 
=       \Sigma_{cm}^{\phantom{(0)}} 
        + \Sigma_{cm}^{\phantom{(0)}} R_{cm}^{(0)} \Sigma_{cm}^{\phantom{(0)}} 
        + \cdots  .
\end{equation}
In either case discussed below it will turn out that only the second-order term
\[
\Sigma_{cm}^{(2)}(i\omega_n) 
= \frac{V^2 \rho T}{N}  \sum_p R_m(i\omega_n + i\nu_p) R_0(i\nu_p)  
\]
is relevant in the universal limit of large $D$. 
The propagators $R_{0,m}$ have the spectral decomposition
\begin{equation}\label{R0m-spectral-decomposition}
R_{0,m}(z) = \int \frac{dx}{ z - x-\lambda} \rho_{0,m}(x) 
\end{equation}
and are centered~\cite{Coleman} around $\lambda$.
After performing Matsubara summations, this gives
\begin{eqnarray}\label{F-2}
D\frac{\partial}{\partial D} \frac{\Delta F^{(2)}}{T} 
&=& \frac{V^2 \rho}{N} D \beta \sum_{m,\sigma=\pm 1} 
        \int\int dx \ dy \ \rho_m(x) \rho_0(y)  \nonumber \\
&&      \times \frac{
        \left[ f(x+\lambda) + b(y+\lambda) \right] 
        \left[  f(\sigma D) - f(x-y) \right]
        }{ \sigma D - x + y } ,
\end{eqnarray}
with $f$ and $b$ the Fermi and Bose functions, respectively.
The coefficient of $e^{-\beta\lambda}$ in Eq.~(\ref{F-2}) for 
large $D$, when combined with Eqs.~(\ref{DF-ef-derivative})
and~(\ref{DF-lambda-derivative}) yields because of
Eq.~(\ref{F-lambda-projection}) the desired scaling
equation~(\ref{F-scaling}).   

In Eq.~(\ref{R-series}) higher-order terms in $\Sigma$ are of
higher order in $e^{-\beta\lambda}$, and vanish relatively to $\Sigma_{cm}$ in
the limit of large $\lambda$. 
By the same arguments as given in Sec.~\ref{Second scaling equation}, one
can show that skeleton diagrams of higher order in $\Sigma_{cm}$ are
irrelevant: 
Higher-order skeleton diagrams in Eq.~(\ref{R-series}) contain at least three
impurity propagators which have to be evaluated at $i\omega_n  = \pm D$ to
make a contribution $\propto  1/D^2$, or they are of higher order in
$e^{-\beta\lambda}$.

\subsection{Conserving slave-boson approach}
\label{Conserving slave-boson approach}

Kroha {\it et al.}~\cite{Kroha/Hirschfeld/Muttalib/Woelfle} developed, for the
Anderson model, a method quite similar to the NCA. 
It is constructed from skeleton diagrams within Coleman's
slave-boson technique.   
Kroha {\it et al.}, however, imposed the constraint only in the average 
\mbox{$\langle  n_f + n_b \rangle_H = 1$}. 
Their method then does not violate the gauge symmetries of the model like the
usual slave-boson mean-field
approach~\cite{Kroha/Hirschfeld/Muttalib/Woelfle}. 
The boundary condition for the free energy is
\begin{equation}\label{Kroha-boundary-conditon}
\frac{\partial}{\partial\lambda} \Delta F = \langle  n_f + n_b \rangle_H = 1 . 
\end{equation}
This includes unphysical states.
Consequently, the conduction electrons acquire a self-energy. 
The first scaling equation~(\ref{DF-scaling-first}) now reads
\begin{equation}\label{DF-scaling-first-Kroha}
\Delta F =  \left( 
T \frac{\partial}{\partial T} 
+ \epsilon_f\frac{\partial}{\partial \epsilon_f} 
+ \rho V^2\frac{\partial}{\partial \rho V^2}
+ D\frac{\partial}{\partial D}  \right)  \Delta F
+ \lambda  .
\end{equation}
In order to obtain any {\it observable}, $\Delta F$ has to be differentiated
once more with respect to an external field.
Hence this equation is the analog to Eq.~(\ref{F-scaling-first}).

The second scaling equation can be derived from Eq.~(\ref{F-2}) in the limit
of large $D$.
First it is summed over $\sigma$.
Only the term $\propto f(\sigma D)$ survive because the propagators $R$ have
a spectral weight centered around a {\it finite} value of $\lambda$, and 
$\sum_\sigma \lim_{D\to\infty} D/(\sigma D - x + y) = 0$.
With
Eqs.~(\ref{DF-ef-derivative}),~(\ref{DF-lambda-derivative}),
and~(\ref{Kroha-boundary-conditon}), one has 
\begin{eqnarray*}
D\frac{\partial}{\partial D} \frac{ \Delta F^{(2)} }{T}
&=&     \frac{-V^2 \rho\beta}{N}  \sum_m
        \int\int dx \ dy \ \rho_m(x) \rho_0(y)                          \\
&&      \phantom{\frac{V^2 \rho}{N}  \beta \sum_m}
        \times\left[ f(x+\lambda) + b(y+\lambda) \right]                \\
&=&     ( 1- 1/N )  V^2\rho 
        \frac{\partial}{\partial \epsilon_f} \frac{ \Delta F^{(2)} }{T}
        - V^2\rho \beta  .
\end{eqnarray*} 
More generally, because 
$\sum_\sigma \lim_{D\to\infty} D/(i \omega _n + \sigma D )  =  0$
in every propagator of Eq.~(\ref{DF-D-derivative-preliminary}) the
Matsubara frequency $i\omega_n$ can be replaced by $\sigma D$ with an
overall factor $f(\sigma D)$.
In the magnetic limit one has $\lambda \gtrsim |\epsilon_f|$ to
enforce~\cite{Kroha/Hirschfeld/Muttalib/Woelfle} both $n_f\approx 1$ and
$n_b\approx 0$. 
Therefore higher order terms in Eq.~(\ref{R-series}) are irrelevant, because
they are either $o(1/D)$, or of higher order in $e^{-\beta\lambda}$.

Hence the approximation also preserves the exact scaling laws.
This explains why their results for the $f$ propagator are so similar to the 
respective NCA results~\cite{review-Bickers}.
Moreover, one can now predict that taking into account more skeleton diagrams
will not alter the picture qualitatively.

\section{Summary}
\label{Summary}

The Anderson-impurity model shows, in a nutshell, the
difficulties when dealing with strong correlations. 
One encounters the same problems as in high-energy physics:
The perturbation theory of observables at energies $T$ in the
coupling constant diverges logarithmically, $\propto  \ln D / T$. 
The limit of zero energy $T$ in solid-state physics corresponds to
the limit of infinite cutoff $D$ in high-energy physics.
If the Hamiltonian can be shown to be renormalizable, it means that there are
finitely many energy scales in the system, at least in the perturbative region.
If there is but one scale, the system can be described by a ``running''
coupling constant $J(D)$, expressed with the help of the $\beta$ function:  
\begin{equation}\label{beta-function}
\beta(J) = \frac{d \ J(D)}{d\ \ln D}   .
\end{equation}
Reducing the cutoff $D$ of the system does not change its physics
as long as the effective coupling $J(D)$ is changed according
to Eq.~(\ref{beta-function}), to keep the energy scale constant.
However, outside the energy regime where the perturbation theory is valid it
has never been shown that such a $\beta$ function exists at all.

In this paper, the variational functional of Luttinger and Ward was used to
prove both the {\it existence} and {\it form} of the $\beta$ function, or,
in other words, the exact energy scales of an impurity-system were determined. 
It turned out that neglecting the vertex corrections (NCA) suffices to proof
universal behavior, and that families of higher-order skeleton diagrams do
{\it not} alter the energy scales.

The method differs therefore from the conventional renormalization-group
approach.
Within that approach, the fixed points of a flow of effective Hamiltonians is
studied perturbatively as the energy scales of the system are varied. 
The crossover in inaccessible by that method.

In comparison to that, the skeleton diagrams can describe the crossover very
well, as was demonstrated.
However, the theory cannot predict the nature of the fixed-point Hamiltonian.

Further work on this subject will concentrate on the question of whether the
NCA can also be justified as a means to solve the effective impurity model onto
which the infinite-dimensional versions of correlated electron systems can be
mapped~\cite{Metzner/Vollhardt,Mueller-Hartmann-d=infinity}.  
In particular one would like to learn whether the periodic Anderson
Hamiltonian exhibits heavy fermion behavior in this limit, 
or if this is only an artifact of the approximations used.
Also, it should be within the reach of the theory to decide which class of
diagrams should be used to describe the problem of two impurities in a metal.

\section*{acknowledgements}

It was a pleasure to discuss the intricacies of diagrammatic approaches
with Tom Schork and Professor V. Zevin. 
I would like to thank Professor P. Fulde and G. Zwicknagl for suggesting that I
investigate the NCA, and for their constant interest in the progress of this
work.

\appendix

\section{Irrelevance of higher-order skeleton diagrams}
\label{Irrelevance of higher-order skeleton diagrams}

The proof uses the spectral decomposition of the ionic propagators $R_f$.
Every skeleton diagram depends explicitly on $D$ only via the
integration boundaries $\pm D$ coming from integration over the
conduction-electron lines, as shown in Eq.~(\ref{NCA equations}). 
For $\alpha = 0, \dots , N$, let $\Sigma_\alpha^{(2n)}R_\alpha$ be such a
skeleton diagram of $2n$th order with $p$ closed fermion loops and $n$ ionic
propagators $R_i$:  
\begin{eqnarray}\label{Skelett-2n}
\lefteqn{ 
\oint \frac{dz}{2 \pi i } e^{-\beta z} \Sigma^{(2n)}_\alpha(z)R_\alpha(z) }
\nonumber \\
&=&     \frac{\rho^n V^{2n}}{N^{n-p}}
        \oint \frac{dz}{2 \pi i } e^{-\beta z} 
        \prod_{j=1}^n \int_{-D}^D f(\epsilon_j) d\epsilon_j
        \prod_{i=1}^n R_i \left (z+\sum^{(i)}_j \epsilon_j \right)  . 
\end{eqnarray}
Here $ \int d\epsilon_j $ denotes the integration over the $j$th
conduction-electron line.   
Furthermore, the term $\sum^{(i)}_j$ in the argument of $R_i$ indicates that
the sum runs over $j$ if the $i$th propagator sits under the
$j$th conduction-electron line, as in Fig.~\ref{skeleton-sixth-order}, the
numbers indicating the $\epsilon_j$.      
The $i$th propagator $R_i$ of the respective skeleton diagram has the
spectral decomposition
\[
R_i(z+\sum^{(i)}_j \epsilon_j) = \int_{-\infty}^\infty d\lambda_i
\rho_i(\lambda_i) \bigg/ \left(
        z + \sum^{(i)}_j \epsilon_j-\lambda_i  \right)  .
\]
For that reason, the line integral gives only contributions from the real
poles 
\[
z = \omega_k = \lambda_k - \sum^{(k)}_j \epsilon_j  .
\]
For real $\omega$, there holds the relation
\[
\Re R_{0,m}(\omega) = \int_{-\infty}^\infty 
\frac{\rho_{0,m}(\lambda)} {\omega - \lambda} d\lambda   ,
\] 
where the Cauchy-principal value of the integral has to be taken. 
Hence the $k$th pole of the line integrals gives
\begin{eqnarray*}
\lefteqn{
\prod_{j=1}^n \int_{-D}^D f(\epsilon_j) d\epsilon_j
\int_{-\infty}^\infty  d\omega \
e^{-\beta (\omega - \sum^{(k)}_j \epsilon_j)} \rho_k(\omega)  
} \\
&& \times \prod_{i\neq k}^n \Re R_i(\omega-\sum^{(k)}_j\epsilon_j + 
\sum^{(i)}_j \epsilon_j)  .
\end{eqnarray*} 
Because of ${\displaystyle f(\epsilon_j) e^{\beta \epsilon_j} =
f(-\epsilon_j) }$, one shifts those variables $\epsilon_j$ to
$-\epsilon_j$, which occur in $\sum^{(k)}$ but not in $\sum^{(i)}$:
\[
\prod_{j=1}^n \int_{-D}^D f(\epsilon_j) d\epsilon_j
\int_{-\infty}^\infty  d\omega \ 
e^{-\beta\omega} \rho_k(\omega) 
\prod_{i\neq k}^n \Re R_i(\omega + \sum^{(i)}_j \epsilon_j)   ,
\]
where now $\sum^{(i)}_j$ runs over $j$ if the propagator $R_i$ sits
under the $j$th conduction-electron line in the respective skeleton diagram
with cyclic permutated vertices.
In such a diagram, the propagator $R_k$ is the outer one.

Exactly {\it here} one uses the fact that only whole families of skeleton
diagrams are considered:
It was just shown that for every family $F$ of $2n$th order skeleton diagrams
their contribution to the line integral in Eq.~(\ref{Upsilon}) is 
\begin{eqnarray*}
\lefteqn{  \text{Tr}_f \oint \frac{dz}{2 \pi i } e^{-\beta z}
        \Sigma_f^{(2n,F)}(z)  R_f(z) = 2n \int d\omega \ 
e^{-\beta\omega}  } \\
&& \times \left[ \rho_0(\omega) \text{Re\,} \Sigma_0^{(2n,F)} (\omega)  
+ \sum_m \rho_m(\omega) \text{Re\,} \Sigma_m^{(2n,F)} (\omega) \right]  
 .  
\end{eqnarray*}
Here, the operator Re is defined to replace {\it every} propagator in its
argument by its real part.  
The factor $2n$ arises because every skeleton diagram occurs in a family
$2n$ times.  
In $\Upsilon$, every skeleton diagram of order $2n$ depends explicitly on $D$
only via the integration boundaries $\pm D$ of its $n$ integrations over the
conduction-electron lines.
The $i$th integral over a conduction-electron line can be written as 
\begin{equation}\label{Lambda1-Lambda2}
\text{Re\,} \Sigma^{(2n,F)}_f(\omega) 
= \left( \int_{-D}^D \right)^{n-1} 
\Lambda_1(\omega) \int_{-D}^D  d\epsilon_i f(\epsilon_i)
\Lambda_2(\omega+\epsilon_i) ,
\end{equation}
where the part $\Lambda_1$ of the respective diagram does not depend on
$\epsilon_i$, that is, its ionic propagators are lying on the left or the
right of the $i$th conduction-electron line. 
The propagators of part $\Lambda_2$ lie under the $i$th conduction-electron
line.  
$ \left( \int \right)^{n-1} $ hints at the other $n-1$ integrations, weighted
with the respective Fermi functions.
One differentiates with respect to $D$ by evaluating the integral over the
$i$th conduction-electron line $\epsilon_i$ at its integration boundaries 
$\pm D$, 
\[
2n \sum_{\sigma = \pm 1} \int d\omega \
e^{-\beta\omega} \rho_{0,m}(\omega) \left[ \Lambda_1(\omega) D f(\sigma D)
\Lambda_2(\omega + \sigma D) \right]   .
\]
The integral is to be weighted with $1/Z_f$ in order to yield the
$D$ derivative of the respective contribution to the free energy. 
Because  $e^{\beta E_0} Z_f$ goes for low temperatures to 1, the spectral
densities $\rho_{0,m}$ are weighted with $e^{\beta (E_0 -\omega)}$. 
For low temperatures $T \ll D$,  this weighted spectral density contributes to
the integral only for frequencies $\omega \lesssim  E_0$.
Because the other integrations over conduction-electron lines $\epsilon_j$ are
weighted with Fermi functions $f(\epsilon_j)$, respectively, 
for low temperatures $T \ll D$ the real part of a propagator 
$\Re R_f(\omega-D+\sum_j \epsilon_j )$
contributes in $\Lambda_2(\omega - D)$ only for frequencies
\[
\omega-D \lesssim  E_0 -D <  - D  .
\]
In this frequency interval one can replace $\Re R_f$ by its naked
counterpart $1/(\omega - H_f - D)$, and hence can estimate its
contribution to $F_f$ from above as $1/D$. 
The term $\propto f(D)$ does not contribute for $T \ll D$.  
It therefore can be estimated as
\[
\frac{1}{Z} D \frac{\partial}{\partial D} \Sigma^{(2n)}_f R_f 
\propto D/D^{m_i}  . 
\]
Hence at least one $m_i$ must equal 1, otherwise the skeleton diagram is
irrelevant. 
However, such a diagram is a skeleton {\it only} if it is of second order;  
otherwise it would have a self-energy insertion. 
This was shown. 
Eq.~(\ref{F-scaling-second}) follows.

\section{NCA at zero temperature}
\label{NCA at zero temperature}

\subsection{NCA differential equations}
\label{NCA differential equations}

The NCA equations~(\ref{NCA equations}) together with the
Dyson equations~(\ref{Dyson}) constitute a self-consistent system of equations
which have been solved numerically~\cite{Bickers/Cox/Wilkins}. 
For $T =  0$ the NCA integral equations~(\ref{NCA equations}) in a
magnetic field are transformed into differential equations by substituting for
the Fermi function a
step function~\cite{Mueller-Hartmann-NCA,KuramotoIII,review-Bickers}, by  
introducing the negative, inverse of the propagators $R_{0,m}$,  
\begin{eqnarray}\label{Y_0,m}
Y_0(z) &=& \Sigma_0^{(2)}(z) - z   ,       \nonumber \\
Y_m(z) &=& \Sigma_m^{(2)}(z) + \epsilon_f - z  ,
\end{eqnarray}
and $Y_m  =  \overline{Y}_m + m g\mu_B h$
for $-j \leq m \leq j  =  (N-1)/2$,
\begin{eqnarray}\label{NCA-Dgl}
\frac{\partial}{\partial\omega} Y_0 &=&  -1 -\frac{\rho V^2}{N}\sum_m
\frac{1}{  \overline{Y}_m + m g\mu_B h  }
\nonumber \\
\frac{\partial}{\partial\omega} \overline{Y}_m &=&  
- 1 -\frac{\rho V^2}{N} Y_0^{-1}  .
\end{eqnarray}
The inverse propagators have the asymptotic forms
\begin{eqnarray}\label{Y0m-asymptotisch}
\overline{Y}_m(\omega) &\approx& - \omega + \epsilon_f   \\
Y_0(\omega) &\approx& - \omega \nonumber 
\end{eqnarray}
for large, negative  $\omega$.
The NCA differential equations have, up to terms of order $O(1/D)$ the integral
\begin{eqnarray}\label{Integral-NCA}
\lefteqn{
\frac{Y_0}{\rho V^2} +  \frac{1}{N} \ln\left| \frac{Y_0}{\rho V^2} \right| 
} \nonumber \\
&=&     \frac{\overline{Y}_m}{\rho V^2} + \frac{1}{N} \sum_m \ln  
        \left| \frac{ \overline{Y}_m+mg\mu_B h }{\rho V^2} \right| 
        - \frac{\epsilon_f^*}{\rho V^2}   \nonumber \\
\epsilon_f^* 
&=&     \epsilon_f + \left( 1-\frac{1}{N} \right) 
        \rho V^2 \ln\frac{D}{\rho V^2}  . 
\end{eqnarray}
The value of $\epsilon_f^*$ follows by inserting Eq.~(\ref{Y0m-asymptotisch}). 
Terms of order $O(1/D)$ are neglected because the universal behavior of the
system is investigated.
The integral of the NCA differential equations contains the energy scales
$\rho V^2$ and $T_K$ via~\cite{Mueller-Hartmann-NCA}
$\ln ( T_K/\rho V^2  ) = \epsilon_f^* / \rho V^2$.
It is a nontrivial task to solve this differential equations numerically
in the universal, magnetic limit
\begin{eqnarray*}
D &\to& \infty  ,  \ \ \epsilon_f \to - \infty \\
\lim_{D\to\infty} \frac{\epsilon_f}{D} &=& 0  , \ \ 
\frac{V^2}{|\epsilon_f|} = J = \text{const.}  .
\end{eqnarray*}

\subsection{What is the ground-state energy of the NCA?}
\label{What is the ground-state energy of the NCA?}

The ground-state energy of the NCA is defined as 
\begin{equation}\label{NCA-E0-definition}
E_0^{\text{NCA}} = \lim_{T\to 0} F_f^{\text{NCA}}  .
\end{equation}
However, one has not yet succeeded in deriving an expression for
$E_0^{\text{NCA}}$ via that route, but merely solves the NCA-differential
equations~(\ref{NCA-Dgl}).
It has been conjectured\cite{review-Bickers} that the lowest, real zeros of
the inverse propagators $Y_{0,m}$ define the NCA ground-state energy. 
This is now proved:
Because of Eq.~(\ref{NCA-E0-definition}) $E_0^{\text{NCA}}$ fulfills 
\[
1  =    \lim_{T\to 0} e^{\beta E_0^{\text{NCA}}}  \int e^{-\beta \omega} 
        \left[ \rho_0^{\text{NCA}} (\omega) 
        + \sum_m \rho_m^{\text{NCA}} (\omega) 
        \right] d\omega   . 
\]
Hence, for $T=0$ there exist the (positive) spectral densities
\begin{equation}\label{rho-hat-definition}
\widehat{\rho}_{0,m}^{\text{NCA}}(\omega) = e^{\beta (E_0^{\text{NCA}}-\omega)}
\rho_{0,m}^{\text{NCA}}(\omega) ,
\end{equation}
and therefore the spectral densities $\rho_{0,m}^{\text{NCA}}$ vanish for
$T=0$ and $\omega < E_0^{\text{NCA}}$. 
This means that the inverse propagators $Y_{0,m}$  are real for those
frequencies.  
In addition, $\widehat{\rho}_{0,m}^{\text{NCA}}$ vanish for $\omega > E_0$,
because $\rho_{0,m}^{\text{NCA}}$ remain finite.  
However, from the existence theorem for solutions of the differential
equations~(\ref{NCA-Dgl}) it follows that there exists at least one value
$\omega_0$ such that the following hold.

(1) $Y_{0,m}$ have a zero in $\omega_0$ and become complex above, if
they are real for large, negative $\omega$ -- which they are in view of
Eq.~(\ref{Y0m-asymptotisch}).
Because it was just shown that the $Y_{0,m}$ are real for 
$\omega < E_0^{\text{NCA}}$, they are real for $\omega < \omega_0$ as
well. 

(2) $\rho_{0,m}^{\text{NCA}}$ vanish for $\omega < \omega_0$,
therefore $\omega_0 \leq E_0^{\text{NCA}}$.  

(3) $\widehat{\rho}_{0,m}^{\text{NCA}}$ vanish for $\omega < \omega_0$,
therefore $\omega_0 \geq E_0^{\text{NCA}}$. 

To sum up:
$\omega_0$ is the lowest common zero for $Y_{0,m}$, and at the same time the
NCA ground-state energy.

\subsection{Parametrization of the NCA ground-state energy}
\label{Parametrization of the NCA ground-state energy}

Hence the well-known expression for the common zero of
$Y_{0.m}$~\cite{Mueller-Hartmann-NCA,KuramotoIII,Inagaki,Kuramoto/Kojima} 
can be used as the NCA ground-state energy and is parametrized in the
following manner:
Define the $W$-function for positive $x$ as 
\begin{equation}\label{W-function-definition}
W(x)\exp\left[ W(x) \right] = x \ \ \text{ for } x \geq 0 ,
\end{equation}
with asymptotic behaviors
\begin{eqnarray}\label{Asymptotik-W}
W(x) &=& x + o(x^2) \ \ \text{ for } x\to 0^+   , \nonumber \\ 
W [ \exp(x) ] 
&=& x - \ln(x) + o(1) \ \ \text{ for } x\to\infty   .
\end{eqnarray}
The integral~(\ref{Integral-NCA}) can be solved for $Y_0$ because both
$Y_0$ and $Y_m$ are positive for $\omega <  E_0^{\text{NCA}}$:
\begin{eqnarray}\label{Y0(Ym)}
\lefteqn{ NV^{-2} Y_0 } \nonumber  \\
&=&     W  \left(  N \exp  \left[ (\overline{Y}_m -
        \epsilon_f^*)\frac{N}{\rho V^2} +  
        \sum_m \ln  \frac{\overline{Y}_m +m g \mu_B h}{\rho V^2} \right]
        \right) . 
\end{eqnarray}
$E_0^{\text{\text{NCA}}}$ can now be parametrized with the help of 
\[
E_0 = D + \int_{-D}^{E_0} d\omega = D + 
\int_{\overline{Y}_m(-D)}^{jg\mu_B h} \frac{d\omega}{d \overline{Y}_m}
d\overline{Y}_m    ,
\]
the NCA differential equations~(\ref{NCA-Dgl}), and Eq.~(\ref{Y0(Ym)}) as
\begin{eqnarray}\label{NCA-E0-magnetischer-Limes}
\lefteqn{ E_0 = \epsilon_f - j g \mu_B h - 
\int_{j g \mu_B h}^{D + \epsilon_f} d\omega  
} \nonumber \\  
&& \times \frac{1}{
        \displaystyle 1+ W 
        \left(  N \exp  \left[ (\omega - \epsilon_f^*)\frac{N}{\rho V^2} + 
        \sum_m \ln  \frac{\omega+m g \mu_B h}{\rho V^2} 
        \right]  \right) } . 
\end{eqnarray}
For vanishing magnetic field there exists another parametrization
because $Y_m$ can be written as a function of $Y_0$ with the help of the 
$W$ function:
\begin{equation}\label{NCA-E0-nichtmagnetischer-Limes}
E_0 =  - \int_0^{D} \frac{d\omega }{\displaystyle
1+   W\left( \exp\left[ (\omega+ \epsilon_f^*)\frac{1}{\rho V^2}   + 
\frac{1}{N} \ln\frac{ \omega}{\rho V^2} \right]  \right) }   .
\end{equation}
The NCA ground-state energy fulfills the equations~(\ref{F-scaling-first})
and~(\ref{F-scaling-second}), as can be checked with the help
of~(\ref{NCA-E0-nichtmagnetischer-Limes}). 
The ground-state energy is {\it not} an universal function because of the
constant term $-\rho V^2$ in Eq.~(\ref{F-scaling-second}).
The ground-state energy up to order $1/N$ and $1/D$ follows
as~\cite{Buch-Hewson}:   
\begin{eqnarray}\label{NCA-E0-1/N-Potenzreihe}
E_0 &=& \epsilon_f 
+       \rho V^2 W \left[ \frac{D}{\rho V^2} \exp\left( \frac{\epsilon_f}
        {\rho V^2}                              \right) \right] \\
&+&     \frac{1}{N}  \int_0^D \frac{W(y)}{[ 1+W(y) ]^3} \ln\frac{x}{D} dx 
        + O(1/D) + O(1/N^2)                     \nonumber       \\
y &=& \frac{D}{\rho V^2} \exp\left( \frac{x+\epsilon_f}{\rho V^2}
        \right)   . \nonumber
\end{eqnarray}

\subsection{An analytical expression for the static magnetic susceptibility of
the NCA} 
\label{An analytical expression for the static magnetic susceptibility of the
NCA}

\subsubsection{Universality}
\label{Universality}

With the help of the Kondo temperature~(\ref{TK}),
Eq.~(\ref{NCA-E0-magnetischer-Limes}) reads, in the magnetic limit
$T_K/\rho V^2 \to 0 $ after the substitution $\omega/ T_K = x $,
\begin{eqnarray}\label{E0-magnetischer-Limes}
E_0 &=& \epsilon_f - j g \mu_B h - T_K
\int_{j g \mu_B h/ T_K }^{(D+\epsilon_f)/T_K } dx \\
&& \times \frac{1} {\displaystyle 1+ W 
\left\{ N \ \exp\left[ \sum_m \ln ( x + \frac{m g \mu_B h}{ T_K })  
\right] \right\} } . \nonumber 
\end{eqnarray}
After differentiating with respect to the magnetic field $h$ the integrand
decays rapidly enough so that one can replace the upper limit of integration
by $\infty$.   
Hence the magnetization scales as a function of $h$ as
\begin{equation}\label{M-Skalierung}
M(h,D,\epsilon_f,V,N) = M \left( \frac{ g\mu_B h }{ T_K } \right)  .
\end{equation}
This shows explicitly that already the skeleton diagrams of second order give
the exact energy scale for the spin degrees of freedom.

\subsubsection{Small magnetic fields}
\label{Small magnetic fields}

Using the parametrization~(\ref{NCA-E0-magnetischer-Limes}), the magnetization
$M(h)  =  -\partial_h E_0$
vanishes for $h=0$ because of $\sum_m m = 0 $. 
The second derivative gives the static magnetic susceptibility for $h=0$.
With the abbreviation~\cite{review-Bickers}
\[
\frac{1}{3}  \mu_j^2 N = (g\mu_B)^2 \sum_m m^2 ,
\] 
from Eq.~(\ref{NCA-E0-magnetischer-Limes}) one has, after substituting $N
Y_0/\rho V^2 = W$, using the NCA differential equations, and finally partially
integrating,
\begin{eqnarray}\label{chi-Darstellung}
\chi(0) &=&     \frac{1}{3} \mu_j^2 \frac{1}{\rho^2 V^4} \int_0^D
\frac{2W(y)+1}{W(y) \left[ 1+W(y) \right]^3} dx         \nonumber       \\
y       &=&     \left( \frac{x}{\rho V^2} \right)^{\frac{1}{N}}  
                \frac{ T_K }{ \rho V^2 } 
                \exp\left(  \frac{x}{\rho V^2} \right)   . 
\end{eqnarray}
In the magnetic limit, $y$ is very small for $x <  |\epsilon_f^*|$ because of
$-\epsilon_f^*  \gg  \rho V^2 $, 
and above that $W > 1$ and the integrand is small.  
Therefore one can replace the $W$ function by its argument, and extend the
integration to $\infty$:
\begin{eqnarray}\label{chi-NCA}
\chi^{\text{NCA}} (0) 
&=&     \frac{1}{3} \mu_j^2 \frac{1 }{T_K} 
        \int_0^\infty e^{-t} t^{-1/N} dt \nonumber \\
&=&     \frac{1}{3} \mu_j^2 \frac{1}{T_K} \Gamma(1-1/N)  .
\end{eqnarray}
The exact result 
\begin{equation}\label{chi-exakt}
\chi(0) = \frac{1}{3} \mu_j^2 \frac{ 1 }{T_K} \frac{1}{\Gamma(1+1/N)} 
\end{equation}
was obtained by fitting the results of the Bethe ansatz to perturbation theory
in the nonmagnetic limit~\cite{Rasul/Hewson}. 
Both results coincide up to order $1/N$, because the NCA contains all diagrams
up to that order. 
This contradicts the claim of Kuramoto and Kojima~\cite{Kuramoto/Kojima} that
the NCA would yield the exact result for $\chi(0)$ in the magnetic limit.

\section{Coqblin-Schrieffer model}
\label{Coqblin-Schriefer model}

\subsection{Variational functional}
\label{Variational functional}

The Schrieffer-Wolff transformation~\cite{review-Bickers} projects -- up to a
constant -- the Anderson model onto the Coqblin-Schrieffer model in the
magnetic limit $\epsilon_f\to -\infty$ and constant
$ J  =  V^2 / |\epsilon_f | $,
\begin{eqnarray}\label{H-CS}
H       &\mapsto& H_{\text{CS}}- 2 \frac{\rho J D}{N} \\
H_{\text{CS}}   &=& 
    \sum_{|\epsilon_p| < D,m} \epsilon_p c^+_{pm} c^{\phantom{+}}_{pm} 
  + \frac{J}{N} \sum_{pq,mn} c^+_{pm} c^{\phantom{+}}_{qn} f_n^+
    f_m^{\phantom{+}}  . \nonumber   
\end{eqnarray}
For that the spectrum of $H$ is shifted at $(-\epsilon_f)$ by shifting the
argument $z$ in $\Upsilon$ at $\epsilon_f$, and the propagators and
self-energies are transformed as~\cite{review-Bickers}
\begin{eqnarray*}
\frac{1}{\epsilon_f} \Sigma_0(z+\epsilon_f) &\rightarrow& 
\Pi(z) , \\
\frac{1}{ 1+  \frac{z}{\epsilon_f} -
\frac{1}{\epsilon_f} \Sigma_0(z+\epsilon_f) } &\rightarrow& 
R_0(z) = \frac{1}{ 1-\Pi(z) } , \\
R_m(z+\epsilon_f) &\rightarrow& 
R_m(z) = \frac{1}{ z-\Sigma_m(z) }  .
\end{eqnarray*}
A $2n$th order diagram carries the prefactor $(-1)^n (\rho J)^n $.
Because $|\epsilon_f| \gg  D, |z| $, the term $z/\epsilon_f$ was
neglected in $\epsilon_f R_0(z + \epsilon_f)$ and such the
charge degrees of freedom of the impurity are projected out.
The variational functional $\Upsilon$ now reads 
\begin{eqnarray}\label{Upsilon-CS}
\Upsilon 
&=&     \beta \oint \frac{dz}{2 \pi i }  e^{-\beta z}   \left\{
        \sum_{m,n} \left( 1-\frac{1}{n} \right) \Sigma_m^{(n)}(z)
        R_m(z)\right. \nonumber  \\ 
&& +    \sum_n \left(  1-\frac{1}{n} \right) \Pi^{(n)}(z)  R_0(z) \nonumber \\
&& +    \ln \left[z- \sum_n \Sigma_m^{(n)}(z) \right]           
+       \left. \ln \left[1- \sum_n \Pi^{(n)}(z) \right]  \right\}  
\end{eqnarray}
The saddle-point property of $\Upsilon$ is shown as Bickers
by~\cite{review-Bickers}.

\subsection{Skeleton diagrams of second order}
\label{Skeleton diagrams of second order-CS}

The analog to the NCA is called the ``self-consistent ladder
approximation''~\cite{review-Bickers}. 
The NCA equations follow from Eq.~(\ref{NCA equations}) after projecting as in
the last paragraph~\cite{review-Bickers}: 
\begin{eqnarray}\label{NCA-CS-model}
\Sigma_m (z) &=& - \frac{\rho J}{N} \int_{-D}^D f(\epsilon)
\frac{1}{ 1- \Pi(z+\epsilon) } d\epsilon   , \\
\Pi(z) &=& - \frac{\rho J}{N} \sum_m  \int_{-D}^D f(\epsilon)
R_m(z+\epsilon) d\epsilon  .  \nonumber 
\end{eqnarray}
The first scaling equation follows as in Sec.~\ref{First scaling equation}:
\begin{equation}\label{F-scaling-first-CS}
F_f = T\frac{\partial }{\partial T} F_f + D\frac{\partial }{\partial D} F_f
                          .
\end{equation}
The asymptotic behavior of the self-energies is estimated from
Eq.~(\ref{NCA-CS-model}) for $|\omega|\ll D$ as
\begin{eqnarray}\label{Sigma-asymptotisch}
\Sigma_m(\omega -D) &=& -\rho J D/N + O(J^2) \nonumber \\
\Pi(\omega -D) &=& O ( J \ \ln D )  .
\end{eqnarray}
Because of Eq.~(\ref{H-CS}), $z$ has to be shifted in every propagator by
$-2\rho J D/N$, to describe the Coqblin-Schrieffer model.  
Therefore 
\begin{eqnarray}\label{Rm-asymptotic}
R_m(\omega -D) &=&   
\frac{1}{1+\rho J / N}  \frac{-1}{D} \ \ \text{for}
|\omega| \ll D ,  \\  
R_0(\omega-D) &=& 1  \ \ \text{for} |\omega| \ll D  . \nonumber 
\end{eqnarray}
Higher orders in $J$ are irrelevant as will be shown below.
It follows within the NCA that 
\[
D\frac{\partial }{\partial D} Z_f 
=       -\beta J \oint \frac{dz}{2 \pi i }
e^{-\beta z} \left(  \frac{R_0(z)}{1+\rho J/N}  -  D R_m(z) \right)    . 
\]
Hence the scaling equation for the impurity part of the free energy is
\begin{equation}\label{F-scaling-second-CS}
D\frac{\partial }{\partial D} F_f^{\text{\text{NCA}}} =
\frac{(\rho J)^2}{1+\rho J/N} 
\frac{\partial }{\partial \rho J} F_f^{\text{\text{NCA}}} 
- \frac{\rho J}{N} D . 
\end{equation}
This equation does not change in the universal limit $J\ll D$ if higher orders
of $J$ in Eq.~(\ref{Rm-asymptotic}) are taken into account.
The following scaling law holds therefore for the impurity part of the free
energy: 
\begin{equation}\label{F-scaling-NCA-CS}
(F_f-E_0)^{\text{\text{NCA}}}(T,D,J) = 
(F_f - E_0)^{\text{\text{NCA}}} \left( \frac{T}{ T_K^{\text{CS}} } \right) 
\end{equation}
with the Kondo temperature of the Coqblin-Schrieffer model~\cite{Buch-Hewson}
\begin{equation}\label{TK-CS}
T_K^{\text{CS}} = D \sqrt[N]{\rho J} \exp [ -1/ \rho J ] . 
\end{equation}
For other observables, one has to couple $H$ to suitable external fields.
In particular, the Kondo resonance is reproduced qualitatively correctly.

\subsection{Skeleton diagrams of higher order}
\label{Skeleton diagrams of higher order-CS}

The two limits $D\to\infty$ and $\epsilon_f\to -\infty$ are 
{\it not} interchangeable, as can be seen from comparing the 
Kondo temperatures of Eqs.~(\ref{TK}) and~(\ref{TK-CS}).
Skeleton diagrams of higher order than two {\it are} relevant in $\Upsilon$.
The reason for that is the asymptotic behavior of $R_0$ which goes to {\it
one} at the cutoff. 
If the diagram in Fig.~\ref{skeleton-sixth-order} is logarithmically
differentiated with respect to $D$, the contribution of the second
conduction-electron line does not vanish for large $D$.
In particular it is not possible to prove now that the
energy scale~(\ref{TK-CS}) is the exact one by considering only
skeleton diagrams of second order.
In fact, the NCA still predicts for $N=1$ a low-energy scale~(\ref{TK-CS})
although there is none.

\subsection{NCA at zero temperature}
\label{NCA at zero temperature-CS}

The derivations are as in Appendix~\ref{NCA at zero temperature}.
The inverse of the pseudopropagator $\Pi$ is defined to be
$Y_0 = 1-\Pi$.
The NCA differential equations are up to terms $\propto  1/D$,
\begin{eqnarray}\label{NCA-Dgl-CS}
\frac{\partial}{\partial\omega} Y_0 &=&  \frac{-\rho J}{N}\sum_m
Y_m^{-1}(\omega)  , \\
\frac{\partial}{\partial\omega} Y_m &=&  -(1-\rho J/N) - \frac{\rho J}{N}
Y_0^{-1}(\omega)   . \nonumber 
\end{eqnarray}
With $Y_m = \overline{Y}_m + m g\mu_B h $ and 
$\widetilde{D} = D(1 - \rho J D/N)$ 
the NCA differential equations have up to terms of order $1/D$ the integral  
\begin{eqnarray}\label{Integral-NCA-CS}
\lefteqn{ Y_0 \left( \frac{1}{\rho J}-\frac{1}{N} \right) 
+  \frac{1}{N} \ln Y_0  
} \nonumber \\
&&   =  \frac{1}{N} \sum_m \ln                  
        \left| \frac{ \overline{Y}_m+mg\mu_B h }{ \widetilde{D} } \right| 
        + \left( \frac{1}{\rho J} - \frac{1}{N} \right)   .  
\end{eqnarray}
The ground-state energy is expressible as 
\[
E_0 = - A + \int_{-A}^{E_0} d\omega  ,
\]
where $A$ is a still arbitrary constant.
If $J \ll A \ll D$ and $\lim_{D\to\infty} A = \infty$, 
the integral~(\ref{Integral-NCA-CS}) can be used as in Appendix~\ref{NCA at
zero temperature} to yield, in the universal limit of small $\rho J$,
\begin{equation}\label{chi-NCA-CS}
\chi^{\text{\text{NCA}}}(0) 
= \frac{1}{3} \mu_j^2 \frac{ 1 }{  T_K^{\text{CS}} } 
\frac{ \Gamma(1-1/N) }{ \sqrt[N]{e} }   .
\end{equation}
which is up to $O(1/N^2)$ identical~\cite{Buch-Hewson} to the result of Rasul
and Hewson, where $\Gamma(1-1/N)$ is replaced by $1/\Gamma(1+1/N)$.




\begin{figure}
\caption{\label{Vertices-Anderson} 
Vertices for the Anderson model. 
A dashed line represents the naked propagator of the occupied
$f^1$ configuration with internal quantum number $m$.
A wavy line represents the naked propagator of the unoccupied
$f^0$ configuration.
A solid line represents the propagator of a conduction electron with internal
quantum number $m$.
Every diagram has a spine of alternating wavy and dashed impurity propagators.
The conduction-electron propagators carry no
self-energy~\protect\cite{review-Bickers}.   
} 
\end{figure}

\begin{figure}
\caption{\label{skeleton-second-order}
Skeleton diagram for $\Sigma_0^{(2)}(R_m(z) )  R_0(z)$.  
The last vertex can be identified with the first because of the trace over the
$f$ configurations in Eq.~(\protect\ref{Upsilon}).
The double, dashed and double, wavy lines represent dressed propagators. 
}   
\end{figure}

\begin{figure}
\caption{\label{skeleton-sixth-order} 
Skeleton diagram of order 6 for $\Sigma_0R_0$
} 
\end{figure}

\end{document}